Piotr Dniestrzański Wroclaw University of Economics
piotr.dniestrzanski@ue.wroc.pl

# GINI INDEX AND ANGLE MEASURE
# AS SPECIAL CASES OF A WIDER FAMILY OF MEASUREMENTS

**Abstract**

In this paper we will show that the Gini coefficient and the introduced measure of angular inequality are special cases of a wider indexed family of measurements. We will discuss the properties of the defined class based, inter alia, on a new elementary characterization of the Gini coefficient.

**Designations**.

For the vector $x \in \mathbb{R}_+^n \setminus \{\mathbf{0}\}$ $zero(x)$ is the number of coordinates of vector $x$ equal 0.

For the vector $x \in \mathbb{R}_+^n \setminus \{\mathbf{0}\}$ $zero(x)$ is the number of coordinates of vector $x$ equal 1.

$\bar{x} = \frac{1}{n}(x_1 + x_2 + ... + x_n)$, $x^p = (x_1^p, x_2^p, ..., x_n^p)$, $\mathbf{1}_n = (1,1,...,1) \in \mathbb{R}^n$.

For the vector $x \in \mathbb{R}_+^n \setminus \{\mathbf{0}\}$ we use the symbol $[x,a]$ to denote the vector $[x,a] = (x_1, x_2, ..., x_n, a)$.

The vectors $x \in \mathbb{R}_+^n \setminus \{\mathbf{0}\}$ and $y \in \mathbb{R}_+^k \setminus \{\mathbf{0}\}$ we use the symbol $[x,y]$ to denote the vector $[x,y] = (x_1 x_2, ..., x_n, y_1, y_2, ..., y_k)$.

## 1. INTRODUCTION

Gini index has been functioning in the social sciences for over a hundred years. It is one of the most popular tools used in assessing inequalities generally. In fact, the Gini index is not so much a measure of inequality but a measure of concentration. To put it another way, the Gini index is a kind of "poverty counter" if we use it to evaluate income inequality. It has a kind of built-in sensitivity to small values. Rejecting this sensitivity it should be considered that, for example vectors $(0,0,1)$ and $(0,1,1)$ have the same degree of inequality. However, the Gini index gives here the values $G(0,0,1) = 2/3$ and $G(0,1,1) = 1/3$. In this case, the index value is simply the percentage of extremely poor in the population. Elementary characterization of Gini coefficient presented in the article (Plata-Pérez et al., 2015)[1] shows that this property is not an accidental consequence of the adopted method of calculating the value of the coefficient but it is its base condition. It would seem that some other measures of inequality are deprived of this sensitivity. Especially if the design of the measure is somehow purely mathematical, technical. For example, if we look at the inequality of vector $x \in \mathbb{R}^n$ as at its angular deviation from a vector with the same coordinates, it can be expected that a measure based on this idea will be more "objective". As we shall see, it is not like that at all. It turns out that the angle measure has similar inclinations as the Gini coefficient. Moreover, in spite of different design, both measures constitute a special case of a broader class of measures.

---

[1] Axiom Standarization (A1) which will be mentioned later on.

## 2. GINI COEFFICIENT - LAST CHARACTERIZATION AND GENERALIZATIONS

There are many different patterns that can be used to calculate the Gini coefficient of the vector $x \in \mathbb{R}_+^n \setminus \{\mathbf{0}\}$. One of its forms is

$$G(x) = \frac{\sum_{i,j=1}^{n} |x_i - x_j|}{2n^2 \bar{x}}, \quad (1)$$

Formula (1) can be easily interpreted and reflects to some extent the idea of this measure.

Presented in the article (Plata-Pérez et al., 2015) mathematical characterization of the Gini coefficient gives new opportunities to study its properties, especially in view of other measures of inequality and disproportionality. The elegant axiomatization is contained in four natural conditions, each of which has a simple interpretation. It is shown that the Gini coefficient is the only measure that satisfies the four axioms:

A1. **Scale Independence**: $I(\lambda x) = \lambda I(x)$ for all $x \in \mathbb{R}_+^n \setminus \{\mathbf{0}\}$ and $\lambda > 0$.

A2. **Symmetry**: $I(x^\theta) = I(x)$ for every permutation $\theta$.

A3. **Standarization**: $I(v^{k+1}) = \frac{k}{n}$, where $v^{k+1} = (\underbrace{0,...,0}_{k}, \underbrace{\frac{1}{n-k},...,\frac{1}{n-k}}_{n-k})$.

A4. **Comonotonne Separability**: $I[\beta x + (1-\beta)z] = \beta I(x) + (1-\beta)I(z)$ for every $\beta \in [0,1]$, every $x$, $z \in \mathbb{R}_+^n \setminus \{\mathbf{0}\}$ that are comonotone[2] and $\sum_{i=1}^{n} x_i = \sum_{i=1}^{n} z_i$.

One of the generalizations of the Gini coefficient was proposed in the work (Gakidou et al., 2000), which defined two families of measures: **individual / mean differences** and **inter individual** differences. The latter measure is similar to that proposed (in the following part of this article) by us. Using the designations adopted in our work, this measure is:

$$IID(\alpha, \beta) = \frac{\sum_{i=1}^{n} \sum_{j=1}^{n} |x_i - x_j|^\alpha}{2n^2 (\bar{x})^\beta}$$

where $\alpha \in (0, \infty)$ and $\beta \in [0,1]$. The results of the research on qualities of these measures are presented in the work (Lai et al., 2008). Other generalization of the Gini coefficient can be found in the article (Weymark, 1981).

---

[2] Authors call the cevtors $x, y \in \mathbb{R}_+^n \setminus \{\mathbf{0}\}$ *comonotone* if for each $i, j \in N$ occurs $(x_i - x_j)(y_i - y_j) \geq 0$.

# 3. ANGLE MEASURES OF INEQUALITY AND DISPROPORTIONALITY.

Gini coefficient is sometimes also used as a measure of disproportionality, for example, in assessing the disproportionate distribution of seats among the political parties or the allocation of seats in the EP between EU Member States. Among many other measures of disproportionality the most natural ones are measures based on the angle between the vectors. One of them is the Salton measure, which vectors $x, y \in \mathbb{R}_+^n \setminus \{0\}$ as the degree of proportionality assigns a value $\cos(x, y)$. Accordingly, a measure of disproportionality would then the value $1 - \cos(x, y)$. Now, as an angular measure of vectors $x, y \in \mathbb{R}_+^n \setminus \{0\}$ disproportionality we accept the value $1 - \cos^2(x, y)$. We denote it with the symbol $A$.

**Def. 1. The angle measure of disproportionality** of vectors $x, y \in \mathbb{R}_+^n \setminus \{0\}$ is what we call the value of $A(x, y) = 1 - \cos^2(x, y)$

Of course there is $A(x, y) = \sin^2(x, y)$. Measures based on the angle between vectors are compared in the literature to measures based on the Pearson correlation coefficient. The article (Egghe and Leydesdorff, 2009) provides a mathematical model of the relationship between these measures. Jones and Furnas (1987) report, in turn, geometric interpretation of this diversity. Salton cosine measure is one of the key measures used, inter alia, in author co-citation analysis. The proposed form of angle measure $A$ turns out to be useful when its version for the subject of inequality measure is compared with Gini coefficient

Koppel and Diskin (2009) showed that the measure $\cos(x, y)$ meets eight natural properties formulated by the two researchers. It is easy to show that the measure $A$ meets all of these properties.

A measure of angle $A$ can be adapted in many ways to the problem of evaluation of vector inequality.

One alternative to the Gini coefficient in the case of estimating the inequality coefficient, based on the angle formed by respective vectors. It is obvious that the vector $x \in \mathbb{R}_+^n \setminus \{0\}$ is even more unequal (by vector inequality one should understand the inequality of its subsequent coordinates), the more it is distant from the vector $\mathbf{1}_n$. We propose to adopt as a measure of inequality of vector $x$ the value $A(x) = A(x, \mathbf{1}_n) = \sin^2(x, \mathbf{1}_n)$.

**Definition. 2. The angle measure of inequality** for vector $x \in \mathbb{R}_+^n \setminus \{0\}$ is called $A(x) = \sin^2(x, \mathbf{1}_n)$

Using the properties of the inner product that translates into a measure $A$:

$A(x) = 1 - \dfrac{(x \circ \mathbf{1}_n)^2}{\|x\|_2 \cdot \|\mathbf{1}_n\|_2}$ where $\| \|_2$ is the norm $L_2$. Therefore, we have:

$$A(x) = 1 - \frac{\left(\sum_{i=1}^{n} x_i\right)^2}{n \sum_{i=1}^{n} x_i^2} \qquad (2)$$

$$A(x) = 1 - \frac{(\bar{x})^2}{\overline{x^2}} \qquad (3)$$

The form (3) is convenient in the use of the accounts, and can be a source of interpretation other than the source one i.e. based on the angle between the $x$ and $\mathbf{1}_n$ vectors.

## 4. GENERALIZATION OF GINI COEFFICIENT

Let us now introduce a new family of inequality measures. Its form will be similar to the one proposed in the article (Gakidou et al., 2000). We will show that to this family belong also the Gini coefficient and the angle measure $A$.

**Definition. 3**. **The measure of inequality** $G_p$ of the $x \in \mathbb{R}_+^n \setminus \{\mathbf{0}\}$ vector is called value $G_p(x) = \dfrac{\sum_{i,j=1}^{n} |x_i - x_j|^p}{2n^2 \overline{x^p}}$, where $p \geq 1$.

For $p \geq 1$ a $G_p$ measure is the Gini coefficient.

## 5. RESULTS

**Proposition 1**. For each $p \geq 1$ measure $G_p$ satisfies axioms A1-A3.
**Proof**. Satisfying axioms A1 and A2 follows directly from the definition. We shall demonstrate ow of the axiom A3 is fulfilled

$$G_p(v^{k+1}) = G_p((n-k)v^{k+1}) = G_p(\underbrace{0,\ldots,0}_{k},\underbrace{1,\ldots,1}_{n-k}) = \dfrac{2k(n-k)}{2n^2 \dfrac{n-k}{n}} = \dfrac{k}{n} \quad \blacksquare$$

Gini coefficient takes values in the range $\left[0, \dfrac{n-1}{n}\right]$. It turns out that this relationship is valid for the whole family of $G_p$ measures.

**Proposition 2.** For each $x \in \mathbb{R}_+^n \setminus \{\mathbf{0}\}$ and $p \geq 1$ is $G_p(x) \in \left[0, \dfrac{n-1}{n}\right]$.

**Proof**. The inequality $G_p(x) \geq 0$ is obvious. We will show that $G_p(x) \leq \dfrac{n-1}{n}$. For each non-negative numbers $a,b$ there occurs the inequality $|a-b|^p \leq a^p + b^p$. Thus we have

$$G_p(x) = \dfrac{\sum_{i,j=1}^{n}|x_i - x_j|^p}{2n\sum_{i=1}^{n}x_i^p} \leq \dfrac{\sum_{\substack{i,j=1\\i\neq j}}^{n}(x_i^p + x_j^p)}{2n\sum_{i=1}^{n}x_i^p} = \dfrac{2(n-1)\sum_{i=1}^{n}x_i^p}{2n\sum_{i=1}^{n}x_i^p} = \dfrac{n-1}{n}.$$

So $G_p(x) \in \left[0, \frac{n-1}{n}\right]$. ∎

The value $\frac{n-1}{n}$ is achieved because the $G_p$ measures satisfy the axiom A3.

**Proposition 3.** For each vector $x \in \mathbb{R}_+^n \setminus \{\mathbf{0}\}$ the equality $G_\infty(x) = \frac{zero(x)}{n}$ holds.

**Proof.** $G_\infty(x) = \lim_{p \to \infty} \frac{\sum_{i,j=1}^n |x_i - x_j|^p}{2n^2 \overline{x^p}} = \lim_{p \to \infty} \frac{\sum_{i,j=1}^n |x_i - x_j|^p}{2n \sum_{i=1}^n x_i^p}$

Without changing the generality[3] we can assume that $\max\{x_1, \ldots, x_n\} = 1$. Then, $\lim_{p \to \infty} |x_i - x_j|^p = 1$ if, and only if, $x_i = 1$ and $x_j = 0$ vice versa. In other cases it is $\lim_{p \to \infty} |x_i - x_j|^p = 0$. Then $\lim_{p \to \infty} \sum_{i,j=1}^n |x_i - x_j|^p$ is equal to twice the product of the number of zeros and number of ones in the $x$ vector coordinates, i.e. $\lim_{p \to \infty} \sum_{i,j=1}^n |x_i - x_j|^p = 2 \cdot zero(x) \cdot one(x)$ Similarly, $\lim_{p \to \infty} \sum_{i=1}^n x_i^p$ is equal to the number of vector $x$ coordinates that equal one, that is $\lim_{p \to \infty} \sum_{i=1}^n x_i^p = one(x)$. So we have

$\lim_{p \to \infty} G_p = \lim_{p \to \infty} \frac{\sum_{i,j=1}^n |x_i - x_j|^p}{2n \sum_{i=1}^n x_i^2} = \frac{2 \cdot zero(x) \cdot one(x)}{2n \cdot one(x)} = \frac{zero(x)}{n}$

It remains to exclude the possibility: $one(x) = 0$. However, this equality does not occur because we have $\max\{x_1, \ldots, x_n\} = 1$. ∎

With Proposition 4 we immediately obtain the well-known feature of the Gini coefficient: $G(x) \geq \frac{zero(x)}{n} = G_\infty(x)$. Which means that the value of the Gini coefficient is not less than the percentage of zero coordinates of the given vector. The feature in the case of the Gini coefficient is easy to determine by referring to geometrical interpretation. Double field between the line of perfect equality and the Lorenz curve is not less than the percentage of zero observations. In the case of $G_p$ measure we do not have such an easy geometric interpretation.

**Proposition 4.** Let $x, y \in \mathbb{R}_+^n \setminus \{\mathbf{0}\}$ If $\sum_{i=1}^n x_i = \sum_{i=1}^n y_i$ and $G_2(x) = G_2(y)$ then for all $a \geq 0$ it is $G_2([x, a]) = G_2([y, a])$.

---

[3] Measures $G_p$ satisfy the axiom A1.

**Proof.** The assumptions conclude that $\sum_{i=1}^{n} x_i^2 = \sum_{i=1}^{n} y_i^2$. Then we have.

$$G_2([x,a]) = 1 - \frac{\left(a + \sum_{i=1}^{n} x_i\right)^2}{(n+1)\left(a^2 + \sum_{i=1}^{n} x_i^2\right)} = 1 - \frac{\left(a + \sum_{i=1}^{n} y_i\right)^2}{(n+1)\left(a^2 + \sum_{i=1}^{n} y_i^2\right)} = G_2([y,a]). \blacksquare$$

The same feature does not hold for measure $G_1$, that is, for the Gini coefficient. It is a fact known from the literature (Cowell, 2011). For example for $x = (1,4,5)$, $y = (2,2,6)$ and $a = 2$ we have $G(x) = G(y) = 0,2667$ and $0,2917 = G([x,a]) \neq G([y,a]) = 0,25$. From Proposition 5 one can draw an immediate conclusion.

**Conclusion**.

Let $x, y \in \mathbb{R}_+^n \setminus \{\mathbf{0}\}$ and $z, t \in \mathbb{R}_+^k \setminus \{\mathbf{0}\}$. If $\sum_{i=1}^{n} x_i = \sum_{i=1}^{n} y_i$, $\sum_{i=1}^{n} z_i = \sum_{i=1}^{n} t_i$, $G_2(x) = G_2(y)$ and $G_2(z) = G_2(t)$ then $G_2([x,z)) = G_2([y,t))$.

**Theorem 1.** Gini coefficient and angle measure are special cases of $G_p$ measures.

**Proof.** It is obvious that $G_1(x) = G(x)$. We will show that $A(x) = G_2(x)$.

$$A(x) = 1 - \frac{\left(\sum_{i=1}^{n} x_i\right)^2}{n\sum_{i=1}^{n} x_i^2} = \frac{n\sum_{i=1}^{n} x_i^2 - \left(\sum_{i=1}^{n} x_i\right)^2}{n\sum_{i=1}^{n} x_i^2} = \frac{n\sum_{i=1}^{n} x_i^2 - \sum_{i=1}^{n} x_i^2 - \sum_{\substack{i,j=i \\ i \neq j}}^{n} x_i x_j}{n\sum_{i=1}^{n} x_i^2} = \frac{(n-1)\sum_{i=1}^{n} x_i^2 - \sum_{\substack{i,j=i \\ i \neq j}}^{n} x_i x_j}{n\sum_{i=1}^{n} x_i^2} =$$

$$= \frac{2(n-1)\sum_{i=1}^{n} x_i^2 - 2\sum_{\substack{i,j=1 \\ i \neq j}}^{n} x_i x_j}{2n^2 \frac{1}{n}\sum_{i=1}^{n} x_i^2} = \frac{\sum_{i,j=1}^{n}(x_i - x_j)^2}{2n^2 \overline{x^2}} = G_2(x). \blacksquare$$

Theorem 1 combines two different measures of inequality. It shows that the Gini coefficient and the $\sin^2$ measure belong to one family of measures. What divides them is value of one parameter. This puts it in a different light two seemingly different measures. Opens up new possibilities for study on the properties of these measures.

**Example**. (Simulation of the speed of convergence of $G_p$ measurement with increasing $p$).

With the increase in the value of $p$ parameter the values of $G_p$ measure approach $G_\infty$ i.e. the measure showing only percentage of zero observations. The speed simulation of this convergence for the vector $x = (1,2,3,4)$ shows that it is quite significant. Received values are $G_1(x) = G(x) = 0,25$, $G_2(x) = 0,1667$, $G_3(x) = 0,115$, $G_{10}(x) = 0,0138$, $G_{20}(x) = 0,0008$, $G_\infty(x) = 0$.

With proposition 4 we have $G_\infty(x) = \frac{zero(x)}{n}$. Let us see how the convergence looks like in the case of the vector in which one of the coordinates is 0, $y = (0,1,2,3)$. For the next values of parameter *p* we have $G_1(y) = G(y) = 0,556$, $G_2(y) = 0,476$, $G_3(y) = 0,426$, $G_{10}(y) = 0,339$, $G_{20}(y) = 0,333$, $G_\infty(y) = 0,25$. The rate of convergence measure $G_p$ is clearly visible in the example of the vector $(1,2)$. We then have $G_p(1,2) = \frac{1}{2} \cdot \frac{1}{2^p + 1}$. We have in this case an exponential convergence.

## 6. SUMMARY

The newly introduced family of measures of inequality includes the Gini coefficient as angular and defined. It has been shown that these two factors are closely related. The new family meets three of the four axioms which uniquely characterize the Gini coefficient. Border measure of this family has a nice interpretation as "poverty counter". At the same time the angle measure at least in one element has an advantage over the Gini coefficient (Proposition 5). Naturally, there appears the question of full axiomatization of the $G_p$ family. Perhaps the A4 axiom can be replaced with its generalization that will cover all $G_p$ measures.